\documentclass[acmtog]{acmart}

\usepackage{color}
\usepackage{float}
\usepackage{subcaption}
\usepackage{url}
\usepackage{wrapfig}
\usepackage{mathtools}
\usepackage[ruled]{algorithm2e} 

\usepackage{xcolor}

\SetCommentSty{mycommfont}
\usepackage{listings}
\usepackage{tabularx} 
\usepackage{colortbl} 
\definecolor{OurTableColor}{RGB}{189,235,252}

\def\mM{M}
\def\mV{\mathcal{V}}

\def\mF{\mathcal{F}}
\newcommand{\mX}{X_{\mM}}
\newcommand{\nV}{n}
\newcommand{\nF}{m}

\newcommand{\mC}{\widetilde{\mM}}
\newcommand{\mcV}{\widetilde{\mV}}
\newcommand{\mcF}{\widetilde{\mF}}

\newcommand{\ncV}{\widetilde{\nV}}
\newcommand{\ncF}{\widetilde{\nF}}

\newcommand{\nn}{\mathcal{F}_\Theta}

\def\textrightarrow{ -> }%

\definecolor{deepfuchsia}{rgb}{0.76, 0.33, 0.76}

\newcommand{\ME}[1]{#1}

\usepackage{array}
\makeatletter
\newcommand{\thickhline}{%
    \noalign {\ifnum 0=`}\fi \hrule height 1pt
    \futurelet \reserved@a \@xhline
}
\newcolumntype{"}{@{\hskip\tabcolsep\vrule width 2pt\hskip\tabcolsep}}
\makeatother

\def\xR{\mathbb{R}}
\newcommand{\tin}{\!\in\!}

\def\mM{M}
\def\mV{\mathcal{V}}
\def\mF{\mathcal{F}}
\renewcommand{\nV}{n}
\renewcommand{\nF}{m}

\copyrightyear{2025}
\acmYear{2025}
\setcopyright{cc}
\setcctype{by-nc-sa}
\acmConference[SIGGRAPH Conference Papers '25]{Special Interest Group on Computer Graphics and Interactive Techniques Conference Conference Papers }{August 10--14, 2025}{Vancouver, BC, Canada}
\acmBooktitle{Special Interest Group on Computer Graphics and Interactive Techniques Conference Conference Papers (SIGGRAPH Conference Papers '25), August 10--14, 2025, Vancouver, BC, Canada}
\acmDOI{10.1145/3721238.3730654}
\acmISBN{979-8-4007-1540-2/2025/08}

\begin{document}

\title{CageNet: A Meta-Framework for Learning on Wild Meshes}

\author{Michal Edelstein}
\orcid{0000-0001-9126-1617}
\affiliation{%
    \institution{Technion - Israel Institute of Technology}
    \city{Haifa}
    \country{Israel}}
\email{edel.michal@gmail.com}

\author{Hsueh-Ti Derek Liu} 
\orcid{0009-0001-1753-4485}
\affiliation{%
    \institution{Roblox \& University of British Columbia}
    \city{Vancouver}
    \country{Canada}}
\email{hsuehtil@gmail.com}

\author{Mirela Ben-Chen} 
\orcid{0000-0002-1732-2327}
\affiliation{%
    \institution{Technion - Israel Institute of Technology}
    \city{Haifa}
    \country{Israel}}
\email{mirela@cs.technion.ac.il}

\begin{abstract}
    Learning on triangle meshes has recently proven to be instrumental to a myriad of tasks, from shape classification, to segmentation, to deformation and animation, to mention just a few. While some of these applications are tackled through neural network architectures which are tailored to the application at hand, many others use generic frameworks for triangle meshes where the only customization required is the modification of the input features and the loss function. Our goal in this paper is to broaden the applicability of these generic frameworks to \emph{``wild'' meshes}, i.e. meshes in-the-wild which often have multiple components, non-manifold elements, disrupted connectivity, or a combination of these. We propose a configurable meta-framework based on the concept of \emph{caged geometry}: Given a mesh, a \emph{cage} is a single component manifold triangle mesh that envelopes it closely. \emph{Generalized barycentric coordinates} map between functions on the cage, and functions on the mesh, allowing us to learn and test on a variety of data, in different applications. We demonstrate this concept by learning segmentation and skinning weights on difficult data, achieving better performance to state of the art techniques on wild meshes. 
\end{abstract}



\begin{teaserfigure}
  \includegraphics[width=\textwidth]{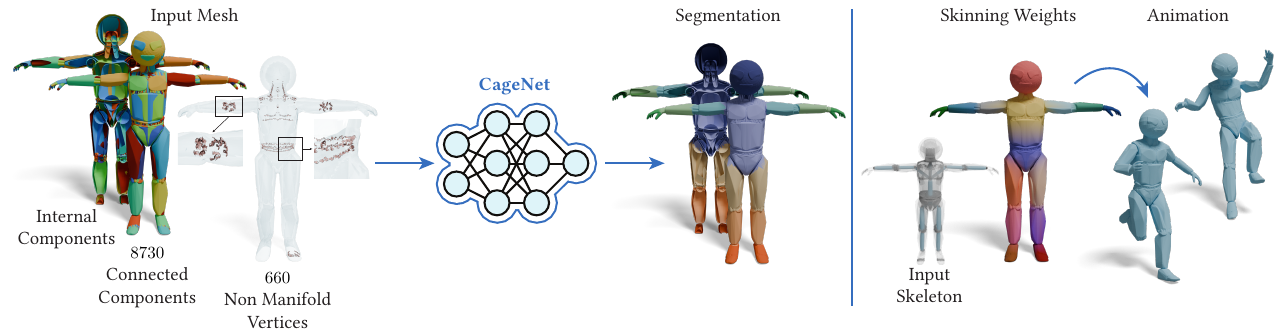}
  \caption{Our CageNet generalizes a segmentation network pre-trained on nearly-manifold inputs to a mesh in the wild (middle) with multiple connected components, non-manifold elements, and internal structures (left). CageNet also enables training on \emph{wild} meshes, such as predicting skinning weights for animation (right).}
  \label{fig:teaser}
\end{teaserfigure}


\maketitle

\section{Introduction}
The impact that neural networks have had on geometry processing is unquestionable. The state-of-the-art methods for applications such as shape classification, segmentation, correspondence, deformation and many others, are all using neural networks, either supervised or unsupervised. These methods are much stronger when they are able to leverage the toolbox of discrete geometry processing, including various discrete differential operators such as the Laplace-Beltrami operator. One such example is DiffusionNet~\cite{SharpACO22}, which is a general architecture that has been successfully applied to many of the previously mentioned applications.

One obstacle to wider adoption of these approaches, is that meshes ``in-the-wild'', e.g. meshes that are generated by artists, are often \emph{inconvenient} from the point of view of geometry processing: they have multiple components, can be non-manifold, and often have non-triangular faces. While geometry processing operators can be potentially generalized to handle these cases, this would inevitably require custom modifications for each artifact, making it not scalable to ``wild'' mesh datasets. Furthermore, multiple-component geometry is specifically difficult to deal with without adding connections that allow information to pass between the components.

We propose CageNet, a meta-framework that fills this gap by enveloping the troublesome geometry with a \emph{cage}, providing a map between the cage and the input geometry using \emph{generalized barycentric coordinates}, and applying a geometry-processing informed neural net on the cage geometry. We use DiffusionNet~\cite{SharpACO22} to demonstrate our framework, due to its wide applicability and easy setup.
We illustrate our framework using two applications: learning mesh segmentation and skinning weights. We achieve comparable performance as the state-of-the-art on a segmentation benchmark consisting of clean mostly-manifold inputs, while obtaining much better generalization to meshes in the wild (see Figure~\ref{fig:teaser}). Compared to networks which are \emph{tailored} to computing skinning weights, our method also achieves the lowest error on wild meshes. Thus, using CageNet one benefits from both worlds: a generic network architecture with minimal setup (defining the input features and the loss), which works out-of-the-box on wild meshes, and achieves the performance of specialized networks. 

\subsection{Related Work}

\begin{figure}[b]
    \centering
    \includegraphics[width=\linewidth]{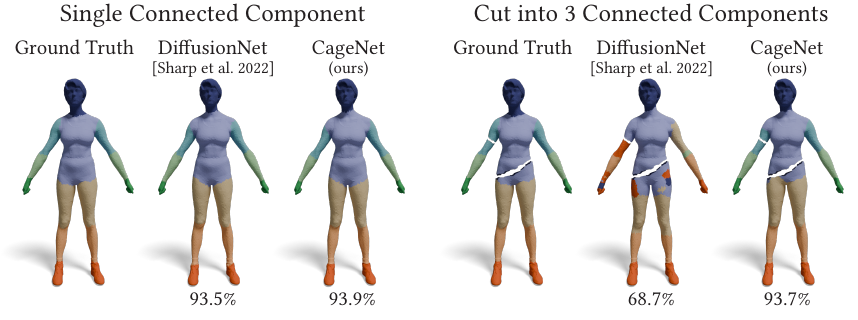}
    \caption{When the input is a single component manifold mesh (left), the existing method \cite{SharpACO22} and our CageNet lead to comparable results. However, when the input contains multiple connected components (right), our method results in better generalization. We show the computed segmentation color coded on the mesh, and the resulting accuracy.}
    \label{fig:seg_cmp_cutmesh}
\end{figure}

\subsubsection{Machine Learning Architectures on Meshes}
Developing machine learning architectures for processing meshes has been an active subject \cite{BronsteinBLSV17}. 
A key element in these architectures is how to generalize common operators, such as message passing or convolution, for signals defined on the surface. 
Several works rely on discrete operators (e.g., convolution) defined on vertices \cite{LimDCK18, GongCBZ19, LahavT20}, edges \cite{HanockaHFGFC19, LiuKCAJ20}, faces \cite{HertzHGC20, HuLGCHMM22}, or a mixture of different mesh elements \cite{MilanoLR0C20}. 
One can also leverage spectral operators, such as the Laplacian, to process scalar \cite{hodgenet, SharpACO22} or vector signals \cite{GaoCKLL24} on surfaces.

Despite their effectiveness, these architectures either fail to process multiple connected mesh components jointly or \emph{non-manifold} elements, which form a majority of the meshes in existing datasets such as ShapeNet \cite{shapenet2015}. 
For instance, the convolution stencil of MeshCNN \cite{HanockaHFGFC19} assumes that each edge only contains two neighboring faces, which is not the case for non-manifold edges. 
Such a limitation poses a significant challenge to deploying existing architectures to mesh data in the wild (see, e.g., Figure~\ref{fig:seg_cmp_cutmesh}).
Even if one considers mesh repair before passing it into a neural network, this often leads to the loss of surface attributes \cite{Hu2018tetwild, HuSWZP20} or to degraded geometric quality \cite{manifoldplus} \ME{(see also the Supplemental).}
This motivates our CageNet to operate directly on these multi-component, non-manifold meshes by parameterizing volumetric signals with manifold cages. 

\subsubsection{Cage-based Geometry Processing}
Cages have been utilized for processing geometry, most notably for shape deformation \cite{stroter2024survey}. Additional classic applications are deformation transfer~\cite{ben2009spatial,chen2010cage}, tracking~\cite{savoye2010cage} and computing ``skinning templates''~\cite{ju2008reusable}. 

More recently, cages were used in learning-based deformation with learned key point handles~\cite{Jakab0M0SK21}, learned cages~\cite{yifan2020neural};
cages have been combined with NeRFs for editing and deformation transfer~\cite{peng2022cagenerf,XuH22}, and have been used to deform Gaussian splatting~\cite{huang2024gsdeformer}. Our method is different from these approaches, since it provides a \emph{general framework} for learning different tasks on wild meshes.
\begin{figure*}
    \centering
    \includegraphics[width=\textwidth]{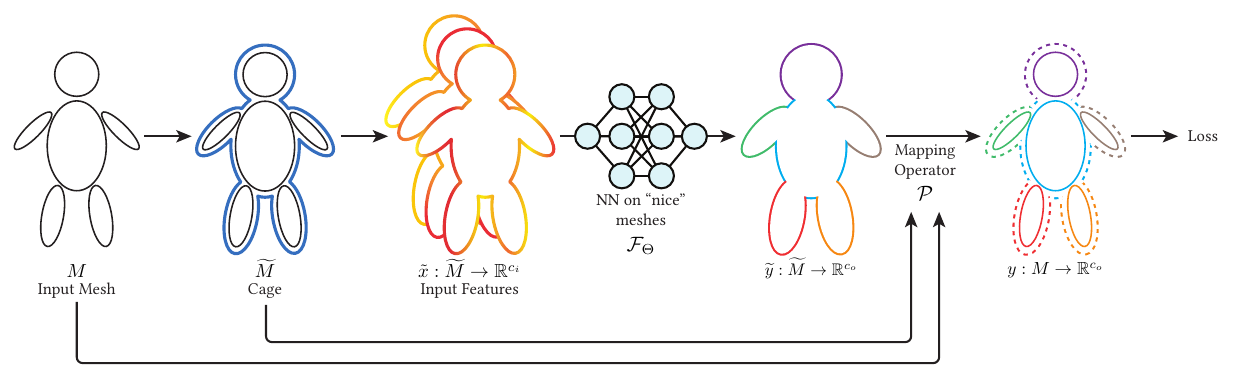}
    \caption{Given an input mesh $\mM$, possibly with multiple components and non-manifold elements, our CageNet constructs a single component, manifold cage $\mC$ and the corresponding mapping operator $\mathcal{P}$ which maps signals from $\mC$ to $\mM$. We then apply a neural network for ``nice'' meshes $\mathcal{F}_\theta$ on the cage $\mC$ (where the input features $\tilde{x}$ are computed). The learned output features $\tilde{y}$ of the cage are projected using $\mathcal{P}$ on the input mesh, yielding the output features $y$ on $\mM$, which are used to compute the loss on $\mM$, where the training data is provided.}
    \label{fig:pipeline} 
\end{figure*}

\subsubsection{Geometry Processing on Wild Meshes}
Another option for handling difficult meshes, is to address each problem separately. Examples include the non-manifold Laplace Beltrami operator~\cite{sharp2020laplacian} (see also citations within for additional examples of non-manifold geometry processing), or adding connectivity to handle meshes with multiple connected components~\cite{garland1997surface,ZhouXTD10}, and other approaches such as repairing meshes~\cite{Hu2018tetwild}. However, adapting an existing network architecture using such approaches would require a custom solution for each type of mesh defect, re-implementing parts of the network and retraining it. Using our framework, it is possible to use off-the-shelf mesh-based architectures and apply them to in-the-wild meshes directly.

\subsection{Contributions}
Our contributions are as follows:
\begin{itemize}
    \item A meta-framework for learning on in-the-wild 3D geometries, such as meshes with interior structures, multiple components, that may be non-manifold.
    \item Using cages and generalized barycentric coordinates to parameterize volumetric functions for neural networks.
    \item Demonstrating the applicability of our framework to shape segmentation and computing skinning weights, achieving better performance on wild meshes than state-of-the-art techniques in a generic framework.
\end{itemize}

\section{Method}
Mesh-based neural network architectures are often restricted to manifold meshes~\cite{HanockaHFGFC19} or have poor performance on meshes with multiple connected components~\cite{SharpACO22}.
CageNet is a neural network architecture that generalizes mesh-based architectures to such \emph{wild meshes}.

Consider a neural network $\nn$ that maps multi-dimensional input features $x$ defined on the vertices (or faces) of the mesh to a set of output features $y$.
Given a non-manifold or multiple component mesh $\mM$, we cannot use $\nn$ directly, either for training or for testing. Instead, we construct a single component, manifold \emph{cage} denoted by $\mC$, that fully encapsulates the model (Sec~\ref{sec:cage}). In addition, we define a \emph{differentiable mapping operator} $\mathcal{P}$ that maps functions on $\mC$ to functions on $\mM$ (Sec~\ref{sec:mapping_op}). We note that the reverse mapping, from $\mM$ to $\mC$ is \emph{not} required. At training time, we apply $\nn$ to the cage $\mC$, and map the resulting output features to $\mM$ using $\mathcal{P}$. Thus, the loss function is applied to features on the input $\mM$, where the training data is provided (Sec.~\ref{sec:cagenet_train}). For testing we again apply $\nn$ to $\mC$, and map the output with $\mathcal{P}$, yielding features on the input mesh $\mM$ (Sec.~\ref{sec:cagenet_test}). This is illustrated in Figure~\ref{fig:pipeline}.

\subsection{The Cage}
\label{sec:cage}
Given a mesh $\mM$, we require a (1) manifold, (2) single component cage $\mC$ such that (3) $\mC$ is geometrically ``close'' to $\mM$ to minimize the loss of detail.
%
Since we apply our method to large datasets, the cage construction must be (4) fully automatic. These are the $4$ necessary requirements for the cage. Additional, nice-to-have properties, are that the cage is (5) topologically equivalent to the input mesh, and (6) that $\mM$ is in the \emph{interior} of $\mC$ for well-behaved cage weights. 

Due to the popularity of cages in geometry processing, there exists a myriad of methods for generating them (see e.g. the recent review by~\citet{stroter2024survey}, Sec. 4.2). Unfortunately, none of the publicly available methods fulfills all our necessary constraints.
Hence, we opted for a basic approach: we compute the unsigned distance field of the input shape, and extract a $\varepsilon$ level set surface with marching cubes (using the implementation by \citet{WangLT22}).
To obtain a single component mesh, we first remove internal components (by computing winding numbers for vertices relative to other components \cite{barill2018fast,gpytoolbox}).
We repeat the process with a larger offset $\varepsilon$ if we obtain multiple disjoint components. The new offset is  $\varepsilon_{new} \!=\! \varepsilon_{old} \!+\! \frac{1}{2}d_{max}$, where $d_{max}$ is the largest distance between two of the disjoint components. This guarantees that the new offset surface has a single component.

\begin{figure}[t]
    \centering
    \includegraphics[width=\linewidth]{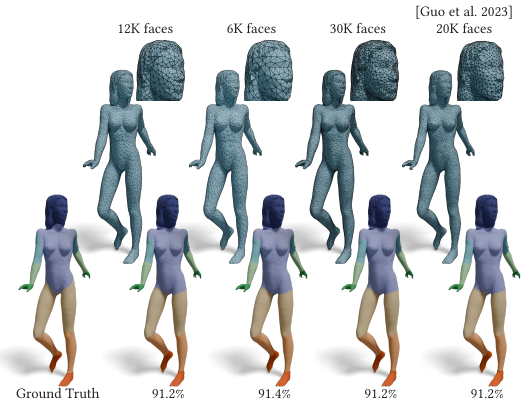}
    \caption{Our method is robust to different cage resolutions, an advantage inherited from DiffusionNet~\cite{SharpACO22}. We show that the segmentation results are comparable (bottom row) even if the cages have different resolutions (top row). We show the accuracy of the segmentation result compared to the ground truth. The three leftmost cages are generated using our approach, whereas the rightmost cage is generated by~\citet{guo2023robust}.}
    \label{fig:seg_cmp_diffcages}
\end{figure}

Finally, we simplify the result using quadric error edge collapse \cite{garland1997surface} as implemented in MeshLab~\cite{cignoni2008meshlab}, to obtain a cage with $\ncF$ faces. We use $\ncF=12K$ 
 for most meshes. \ME{If the result does not encapsulate the input mesh (which happens rarely), and an encapsulating cage is required, we} add more faces
 by simplifying to a larger number of faces, up to $\ncF=24K$.
 Figure~\ref{fig:seg_cmp_diffcages} shows that the process is not sensitive to the exact number $\ncF$, yielding very similar results at test time for different $\ncF$ values.

We emphasize that the process is fully automatic, and results in a single component manifold cage that encapsulates the input mesh, thus fulfilling our requirements. This does not, however, guarantee that the cage is topologically equivalent to the input mesh. Figure~\ref{fig:seg_cageoffset} shows a few resulting cages for different offsets $\varepsilon$. Note that for a large offset, the hands connect to the body, leading to a high genus mesh. We optionally compute multiple cages for each mesh to augment the data, see the segmentation experiment in Sec.~\ref{sec:exp_seg}.

\subsection{The Mapping Operator $\mathcal{P}$}
\label{sec:mapping_op}
The mapping operator $\mathcal{P}$ takes as input the mesh $\mM = (\mV, \mF)$, the cage $\mC = (\mcV, \mcF)$ and a function on the cage $\tilde{y} : \mcV \to \xR$, and computes a function on the input mesh $y: \mV \to \xR$. We opted for a very simple solution, using a linear map computed using \emph{generalized barycentric coordinates}. Given a point $p \tin \mM$, its generalized barycentric coordinates with respect to the cage $\mC$ are $\{\lambda_1(p), \lambda_2(p), ..., \lambda_{\ncV}(p)\} \tin \xR^{\ncV}$, such that $\sum_j\lambda_j = 1$~\cite[Sec 2.1]{stroter2024survey}. Here, $\ncV = |\mcV|$.
The mapping operator $\mathcal{P}$ is thus defined as:
\begin{equation}
\label{eq:mapping_op_p}
    \mathcal{P}(\mM, \mC, \tilde{y}) = C_{\mC}(\mX) \tilde{y},
\end{equation}
where $\mX \tin \xR^{\nV\times3}$ are the coordinates of the vertices of the input mesh $\mM$, with $\nV=|\mV|$. Further, $C_{\mC}$ is a matrix of size $\nV \times \ncV$, whose $(i,j)$-th entry is $\lambda_j(p_i)$ where $p_i \tin \xR^3$ is the embedding of the vertex $v_i \tin \mV$. Hence, $y=\mathcal{P}(\mM, \mC, \tilde{y})$ contains weighted averages of the values of $\tilde{y}$, where the weights are given by the  generalized barycentric coordinates.

Different coordinate functions have different expressivity, and computational complexity. E.g., the Mean Value Coordinates (MVC) by Floater et al.~\shortcite{Floater03} have a closed form expression, \ME{do not require an encapsulating cage} and are very efficient to compute. On the other hand, Harmonic~\cite{JoshiMDGS07} and biharmonic~\cite{JacobsonBPS11} coordinates \ME{require an encapsulating cage}, do not have a closed form, and require solving a PDE based optimization problem, leading to much higher computational costs. In practice, for PDE based coordinates, we create a tetrahedral mesh of the cage \cite{HuSWZP20}, solve the PDE using FEM, and linearly interpolate the coordinates from the vertices of each tetrahedron to its interior.

In terms of expressivity, MVC and biharmonic coordinates are similar, whereas harmonic coordinates struggle to represent encapsulated structures with different function values. Since a harmonic function is determined by its boundary values, this is to be expected. 
\begin{figure}[t]
    \centering
    \includegraphics[width=\linewidth]{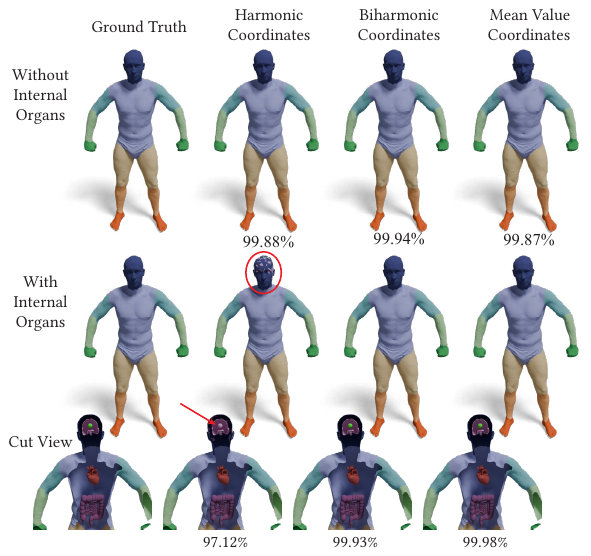}
    \caption{The choice of coordinates influences the generalization of CageNet. When the input is a single manifold mesh without interior structure (top row), CageNet can overfit the segmentation perfectly using all three coordinate choices. But when the input contains interior components (bottom rows), the harmonic coordinates fail near the brain region (second column), while the biharmonic (third column) and the mean value coordinates (fourth column) still lead to perfect predictions.}
    \label{fig:coor_cmp}
\end{figure}

Figure~\ref{fig:coor_cmp} compares these three types of generalized barycentric coordinates for representing a segmentation function. We show the ground truth color coded segmentation, as well as the representation obtained by overfitting CageNet to this function (see Sec~\ref{sec:cagenet_train}), using the different coordinates. For representing the segmentation of a human (top row), all coordinates are equally good. For a shape with ``internal organs'' (bottom rows), which are classified differently by the segmentation function, MVC and biharmonic coordinates perform significantly better, while MVC is much faster to compute. Thus, in all our experiments we use MVC.

\subsection{CageNet Training}
\label{sec:cagenet_train}
We first compute all the cages and their corresponding generalized barycentric coordinates for all the input meshes. 

In our experiments, we use DiffusionNet~\cite{SharpACO22} as the underlying network due to its robustness to cage tessellation. Our framework, however, is general, and other mesh-based networks could be used instead. 

We use the default DiffusionNet input features, which are computed on the input cages. We elaborate on the features when describing the specific experiments in Section~\ref{sec:experiments}. We emphasize that the loss is computed \emph{on the input data} directly, by mapping the cage output features computed by the network to the input meshes using the mapping operator $\mathcal{P}$. When the labeling data is provided on the edges or the faces (for example, for segmentation), we first map from the cage to the input mesh vertices, then average to the edges or faces, and apply the last activation as appropriate.

\subsection{CageNet Testing}
\label{sec:cagenet_test}
The cage and barycentric coordinates are computed on the input mesh. Then the network is tested as usual, with the output signal again mapped to the source mesh using the mapping operator $\mathcal{P}$. 

CageNet can generalize from training on ``nice'' meshes, to testing on ``wild'' meshes. For example, in Section~\ref{sec:exp_seg} we train CageNet on the human segmentation benchmark~\cite{MaronGATDYKL17}, where all the models are single component meshes. In Figure~\ref{fig:teaser} we test the trained network on a model which is non-manifold, has internal structures and multiple components. Indeed, the network generalizes well to this type of ``inconvenient'' (though abundant in the wild) meshes. 

Figure~\ref{fig:seg_cmp_diffcages} demonstrates that the method is not sensitive at test time to the cage resolution. Specifically, we test the segmentation network trained in Sec~\ref{sec:exp_seg} using cages with different resolutions generated with our method, as well as a cage generated by the method of~\citet{guo2023robust}. We note that the segmentation results are very similar, regardless of the cage used.

\section{Experimental Results}
\label{sec:experiments}
We apply CageNet to two applications: segmentation and computation of skinning weights. We used the authors' implementation of DiffusionNet~\cite{SharpACO22}, and implemented CageNet in PyTorch \cite{PaszkeGMLBCKLGA19}. All training was done on a single machine with NVIDIA RTX A4500 GPU (20GB GPU memory) and 128GB system memory. We elaborate on the training data, input features, loss, hyper-parameters and evaluation metrics for each application separately.

\begin{figure}[b]
    \centering
    \includegraphics[width=\linewidth]{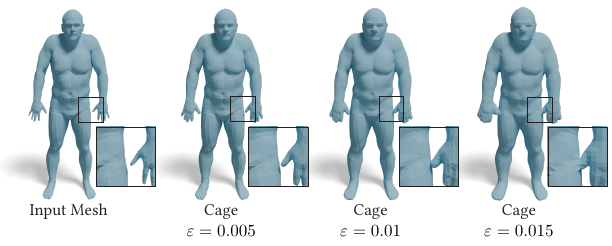}
    \caption{We augment the training data by generating multiple cages with different offset values. This increases the robustness of CageNet against topological noise as the cage topology may change with different offsets, such as the hand region in this example. 
    }
    \label{fig:seg_cageoffset}
\end{figure}
\section{Human segmentation}
\label{sec:exp_seg}
Semantic segmentation \cite{gao2023large} is one of the most basic shape processing tasks, and as such is addressed by many different methods. We focus here on the \emph{Human Segmentation} dataset by \citet{MaronGATDYKL17}, due to its popularity as a benchmark.

\subsection{Implementation Details}

\subsubsection{Input features and hyper-parameters}
We use a very similar setup as DiffusionNet's original segmentation setup, using HKS features. The only different hyper-parameters are the network width (64 instead of 128), and the initial learning rate (0.0005 instead of 0.001). We provide the full list of hyper-parameters in the supplemental material for completeness.

\subsubsection{Loss}
We apply a cross-entropy loss on vertex labels averaged to the faces (as does DiffusionNet). In our case, to obtain the vertex labels we use the operator $\mathcal{P}$ to map the cage functions to functions on the vertices of the input mesh.

\subsubsection{Evaluation Metrics}
As is standard, we use the segmentation accuracy as the evaluation metric.

\subsection{Results}
In Table~\ref{tab:seg_cmp} we show that our method achieves the same results as DiffusionNet on \emph{nice} single-component manifold meshes. Indeed, since our method is based on DiffusionNet, it cannot be expected to surpass it. In contrast, when testing on a broken mesh as in Figure~\ref{fig:seg_cmp_cutmesh}, our method generalizes much better than DiffusionNet. Here, we remove some faces from one of the meshes in the dataset to generate a mesh with multiple connected components. This allows us to compute the accuracy, since we have the ground truth data for the remaining faces. \ME{We also show our robustness to wild meshes by testing on a \emph{"souped"} dataset. We split each model in the test dataset into a triangle soup and randomly inverted half of the face normals. We obtain $91.7\%$ accuracy, the same as for the clean dataset.}

As another experiment, in Figure~\ref{fig:seg_cmp_roblox} we test on \emph{wild mesh} inputs, which have multiple connected components. Here as well our method leads to better generalization than the baselines. Figure~\ref{fig:teaser} shows another example, where we test on a mesh which has multiple components, interior parts and non-manifold elements, and achieve good generalization. For the wild meshes ground truth is not available, so we do not compute accuracy. A qualitative comparison is possible with the ground truth segmentation of one of the meshes from the training dataset, shown on the top left. \ME{We show additional results on wild meshes in the Supplemental material.}
\begin{figure}[t]
    \centering
    \includegraphics[width=\linewidth]{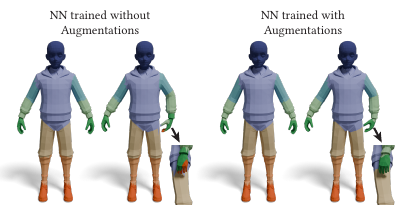}
    \caption{We compare the segmentation results on a wild mesh that does not belong to the training dataset, when the network is trained without (left) and with (right)  cage offset augmentation (see Sec.~\ref{sec:offset_augmentation}).
    In each pair, the mesh on the right has problematic regions (the left hand) which leads to different cage topologies for different offsets, whereas the mesh on the left does not. 
    The mesh on the left yields good segmentation results with or without augmentation (see e.g. the ground truth segmentation in Fig.~\ref{fig:seg_cmp_diffcages}). However, the mesh on the right leads to erroneous segmentation on the left hand without augmentation, which is resolved when adding the augmentation.}
    \label{fig:seg_cmp_topo}
\end{figure}

\subsection{Cage offset augmentation}
\label{sec:offset_augmentation}
For each training mesh we generate cages at offsets $0.005,0.01, 0.015$, and $0.02$ (models are normalized in the unit box), see Figure~\ref{fig:seg_cageoffset}. At each epoch, for each mesh, we randomly choose one of the generated cages. For testing we use the smallest training offset. \ME{Note that while each additional cage increases preprocessing time during training, it does not affect the training loop time. In addition, it does not affect inference time, as testing is performed with a single cage offset.} Figure~\ref{fig:seg_cmp_topo} shows an example of testing using a net that was trained without (left) and with (right) cage offset augmentation. Note the correction obtained for the segmentation of the left hand of the model when the augmentation is used. 
The results on the human segmentation dataset without the augmentation are given in Table.~\ref{tab:seg_cmp}. Indeed, the accuracy without the augmentation is lower.  

\begin{table}
    \centering
    {\smaller
    \caption{We report the segmentation results on the manifold mesh dataset \cite{MaronGATDYKL17}. Our CageNet does not deteriorate the performance of existing methods on clean meshes, getting the same accuracy ($\%$) as the DiffusionNet \cite{SharpACO22} where our method is based on. In contrast, we can achieve better generalization to wild inputs in Figure~\ref{fig:seg_cmp_topo}.}
    \label{tab:seg_cmp}
            \begin{tabular}{@{\hskip 0pt}l@{\hskip 2pt}r@{\hskip 6pt}||@{\hskip 2pt}l@{\hskip 2pt}r@{\hskip 0pt}}
            \hline   
            Method & Acc. & Method & Acc. \\
            \hline
            GCNN \cite{masci2015geodesic} & 86.4 & MeshWalker [Lahav et al. \citeyear{LahavT20}] & 92.7 \\ 
            ACNN \cite{boscaini2016learning} & 83.7 & CGConv \cite{yang2021continuous} & 89.9 \\
            Toric Cover \cite{MaronGATDYKL17} & 88.0 & FC \cite{mitchel2021field} & 92.5 \\
            PointNet++ \cite{qi2017pointnet} & 90.8 & DiffusionNet \cite{SharpACO22} & 91.7 \\
            MDGCNN [Poulenard et al. \citeyear{poulenard2018multi}] & 88.6 & MeshMAE \cite{liang2022meshmae} & 90.0 \\ 
            DGCNN \cite{wang2019dynamic} & 89.7 & SubdivNet \cite{HuLGCHMM22} & 93.0 \\
            SNGC \cite{haim2019surface} & 91.0 & SieveNet \cite{yuan2023sievenet} & 93.2 \\
            PFCNN \cite{yang2020pfcnn} & 91.5 & CageNet (ours) w/o aug. & 90.4 \\
            HSN \cite{wiersma2020cnns} & 91.1 & CageNet (ours) & 91.7 \\
            PD-MeshNet \cite{MilanoLR0C20} & 86.9  & \ME{CageNet (ours) - \emph{"Souped"} dataset} & 91.7 \\
            \hline   
        \end{tabular}
    \vspace{-.1in}
    }
\end{table}

\section{Skinning Weights}
\label{sec:exp_skinning}

Skinning is one of the most popular methods for real-time shape deformation \cite{skinningcourse2014}. A core ingredient is the creation of \emph{skinning weights} to bind vertices with skeletons to drive deformation. Some existing techniques offer a semi-automatic interface to aid in the creation \cite{BangL18, MaLZ24}. Some rely on fully automatic approaches to optimize for smoothed and localized skinning weights based purely on geometric information \cite{JacobsonBPS11, Wang21weights, DionneL13}. However, these techniques may struggle to create weights that are aligned with, for instance, semantics. This issue has been motivating the development of several learning-based techniques, including \cite{liu2019neuroskinning, XuZKLS20, Mosella-Montoro22, MaZ23, PanHMWLSWJ21, ouyang2020autoskin}, to capture semantic information by learning from data, using neural nets \emph{tailored} to this problem. Our approach, on the other hand, is, to our best knowledge, the first to treat this problem (albeit with a known constant skeleton) in a general mesh-based network.

We evaluate the performance on skinning weight generation on a dataset of 753 artist-created biped characters (see supplemental, Figure 1) obtained from the Roblox platform, where we hold out 75 meshes as the test set. Meshes in the dataset share the same pose, orientation and skeleton topology.

\subsection{Implementation Details}
\begin{figure}[b]
    \centering
    \includegraphics[width=\linewidth]{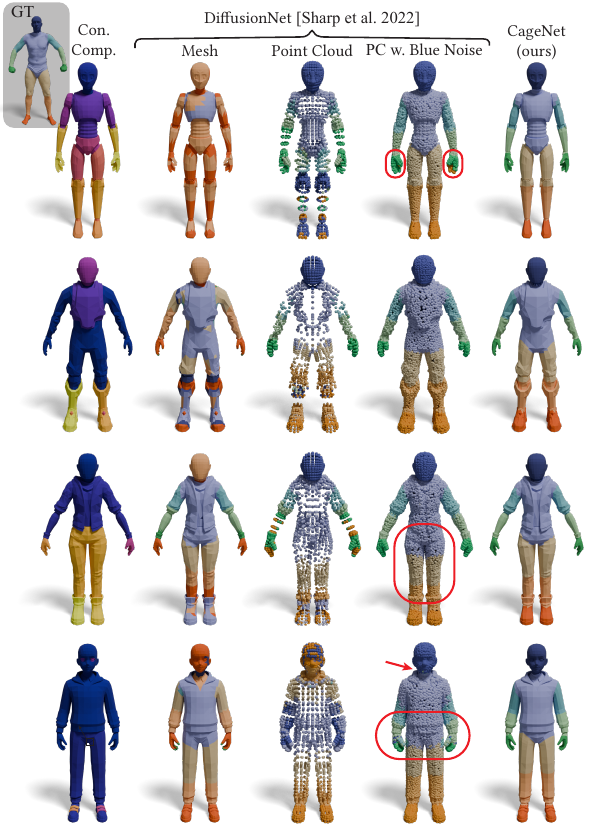}
    \caption{Segmentation results on wild biped characters, consisting of multiple connected components visualized using component-wise colors (first column). DiffusionNet \cite{SharpACO22} trained only on a single component mostly-manifold mesh dataset \cite{MaronGATDYKL17} fails to generalize, regardless of using mesh connectivity (second column), vertex point cloud connectivity (third column), or uniformly sampled \cite{Yuksel2015} point connectivity (fourth column). In contrast, our CageNet (fifth column) generalizes to wild characters to produce desired segmentation results. These in-the-wild models do not have ground truth segmentation. The ground truth on one of the meshes from the training dataset is shown for comparison on the top left.}
    \label{fig:seg_cmp_roblox}
\end{figure}

\subsubsection{Input features and hyper-parameters}
We use a cage offset of $\epsilon = 0.01$. The cage vertex locations and the \emph{volumetric geodesic distance} from the cage vertices to the skeleton's bones are the input features to our network, as suggested in \citet{XuZKLS20}.
Our network uses DiffusionNet's default hyper-parameters, with a 128-width. After obtaining output features on the cage, we use the cage coordinates to map the results to the input mesh. As skinning weights need to be positive and sum up to $1$, we apply the softmax activation to the weights at each vertex to satisfy these constraints.

\subsubsection{Loss}
Our loss function consists of a KL divergence term, an $L_p$ loss to encourage sparsity, and a symmetry loss to encourage symmetric weights for symmetric meshes. Thus our loss is:
\begin{align}
\mathcal{L} = \text{KL}(\hat{s}, s) + \lambda_p L_p (\hat{s}) + \lambda_{\text{sym}} \text{Sym}(\hat{s}, s),
\end{align}
where $s$ are the ground truth skinning weights, $\hat{s}$ are the predicted skinning weights and $\lambda_p$ and $\lambda_{\text{sym}}$ balance the different terms. We use $p=0.3, \lambda_p=0.1, \lambda_{\text{sym}}=0.05$.

The skinning weights per vertex are effectively a distribution with respect to the bones (since they are positive and sum to 1). Hence we use the KL divergence to  measure the error of the predicted weights w.r.t. ground truth, as in~\cite{liu2019neuroskinning}. We compute the KL divergence per vertex, and then average over the vertices.   

The skinning weights are expected to be \emph{sparse}, thus each vertex should be affected by a small number of bones. The $L_p$ loss, with $0 < p \leq 1$ is used in classic optimization methods to encourage sparsity~\cite{bruckstein2009sparse}. While $p=1$ is a common choice \cite{tibshirani1996regression}, it is known that smaller $p$ values are more beneficial for sparsity. We find that $p=0.3$ leads to the best results.
The $L_p$-norm of the skinning weights is computed per vertex, and then averaged over the vertices.

The symmetry loss is designed to improve the symmetry of the predicted weights in areas where both the mesh and the ground truth weights are symmetric. Hence, it measures the symmetry error of the predicted weights in these regions. For simplicity, we consider only extrinsic symmetry with respect to the $yz$ plane, which is compatible with our  consistently oriented dataset.

\subsubsection{Evaluation Metrics}
We use a few evaluation metrics, following~\citet{XuZKLS20}.
(1) The average $L_1$ loss compares the predicted and ground truth weights in the $L_1$ norm. (2) Precision and recall measure the success of finding the \emph{influential} (weight larger than $0.0001$) bones of a vertex, and the $F_1$ score is the harmonic mean of precision and recall, representing both in one metric. (3) We measure the normalized vertex distance between an animation frame generated by the predicted v.s. the ground truth weights, and report the average and the maximum.
The vertex distances are evaluated on 8 artist-created animations (see the supplementary video).

\subsection{Results}
\subsubsection{Quantitative Results}
We evaluate our method by comparing the predicted skinning weights against different baselines, including BBW~\cite{JacobsonBPS11}, GeoVoxel~\cite{DionneL13}, and RigNet~\cite{XuZKLS20}.
For BBW, we use the authors implementation, where we first tetrahedralize the mesh using fTetWild \cite{HuSWZP20}. For GeoVoxel, we use Maya’s implementation \cite{maya2024}. We show the results both with the default parameters, and with the recommended parameters in \cite{XuZKLS20}.
For both of these methods, we remove meshes with joints that are outside the body, which are about $5\%$ of the test set.

RigNet and our CageNet are trained on 90\% randomly selected meshes from our dataset (see Supplementary Figure 1) and evaluated on the test set.
We present our results using our predicted skinning (``Ours'') and with sparsity enforced by thresholding values below 0.2 to zero (``Ours-S'').

The results are reported in Table \ref{tab:sw_cmp}.
We note that for the average $L_1$ distance, which measures the error of the predicted weights, our method achieves the best results (with and without enforcing sparsity). The statistical $F_1$ score, which combines both the precision and recall of finding the influential bones, is also the best for our approach. As expected, enforcing sparsity by thresholding increases precision and reduces recall, however the combined $F_1$ measure shows that the trade-off is reasonable and not much is lost. In terms of vertex distance of the generated animation frames, we achieve the best average distance (tied with RigNet). The best maximal distance is achieved by BBW, with the rest of the methods achieving mostly similar results. BBW wins on this metric, as it is the only method that is not affected by regions of the mesh that are close in Euclidean distance but far in geodesic distance. 

To summarize, our method shows very good results, comparable or surpassing existing techniques. The closest to our performance is RigNet, which is \emph{tailored} for computing skinning weights (and thus for example can also handle different skeleton structures), whereas we achieve these results using a \emph{generic} network.

\begin{figure}[b]
    \centering
    \includegraphics[width=\linewidth]{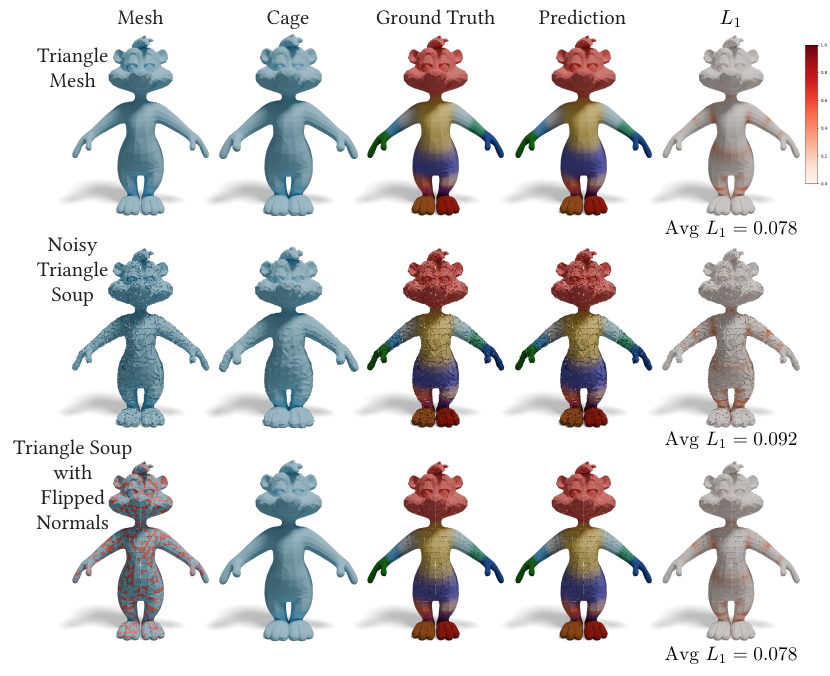}
    \caption{Generalization to triangle soups. We train on a single-component triangle mesh (top) and test on a  triangle soup with vertex noise (middle) and with flipped triangles (bottom). Note that our method leads to the desired skinning weight prediction (4-th column) with low error (5-th column).}
    \label{fig:sw_trisoup}
\end{figure}

\begin{figure*}
    \centering
    \includegraphics[width=\linewidth]{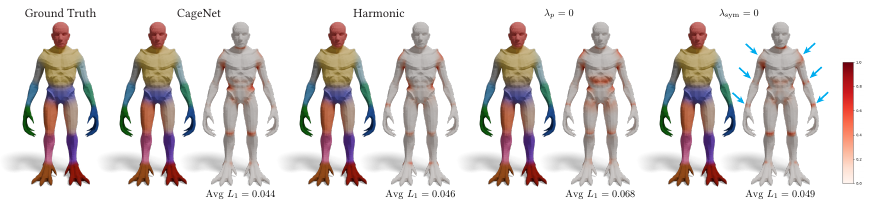}
    \caption{Our ablation study on the skinning weight experiment shows that our choice of coordinates and incorporating the Lp and the symmetry loss terms lead to more accurate predictions. Specifically, removing the $L_p$ loss increases the error significantly, since weights are no longer sparse. Removing the symmetry loss leads to non-symmetric weights (visible in the non-symmetric error) in addition to a higher error term.}
    \label{fig:sw_ablation}
\end{figure*}

\subsubsection{Qualitative Results}
We show qualitative results in Figure~\ref{fig:sw_cmp_test}. We show for each model the rest pose and skeleton (left most column), and the ground truth weights. The weights are visualized by setting a different color for each bone, and combining these colors using the weights. In addition, we show the ground truth for one animation frame from a set of user-generated animations. The animation frame is computed using Linear Blend Skinning (LBS)~\cite{skinningcourse2014} with the ground truth skinning weights. For each method we show the computed skinning weights, and the corresponding animation frame computed with LBS and the computed weights. The animation frame is color-coded with the pointwise normalized vertex displacement error. Note that in all cases our error is considerably lower than the other methods, with RigNet arguably as the closest competitor. We provide the full animation sequence in the supplemental video \ME{and the average normalized vertex error per frame in the Supplemental}.

In addition to baseline comparisons, in Figure~\ref{fig:sw_trisoup} we demonstrate our resilience to multi-component meshes by decomposing a mesh into a triangle soup (second row), as well as adding vertex noise to the triangle soup (third row), and flipping randomly the orientation of some of the faces (fourth row). For each case we show the mesh, the cage, the ground truth weights, predicted weights, and color coded $L_1$ pointwise error, as well as the average error. We note that for both types of noise the predicted result is very similar to the predicted result on the original mesh.

In Figure~\ref{fig:sw_cmp_ai} we apply our method to wild AI meshes generated by \cite{xiang2024structured} and Meshy, and compare it with the baselines. These meshes do not have ground truth, hence we show only the resulting weights, and a corresponding animation frame. The weights can be qualitatively compared to the ground truth weights of the Roblox dataset in Figure~\ref{fig:sw_cmp_test}, since we used the same skeleton (which was manually posed to fit the generated mesh). We note that our results are most similar semantically to the ground truth weights. While RigNet leads to good results as well, artifacts are visible for all three models. See  the video for the animation sequence.

\ME{We also show additional results on wild meshes, and an experiment demonstrating the cage topology effect on the results in the Supplementary.}

\begin{table}[b]
    \centering
    {\small
    \caption{We compare our predicted skinning weights vs. existing baselines: bounded biharmonic weights (BBW \cite{JacobsonBPS11}), GeoVoxel (\cite{DionneL13} GV with default parameters, GV* with parameters recommended by \citet{XuZKLS20}), and RigNet \cite{XuZKLS20}. 
    We report the following metrics: average $L_1$ error ($\downarrow$), Precision ($\uparrow$), Recall ($\uparrow$), F1-score ($\uparrow$), average vertex distance ($\downarrow$), and maximum vertex distance ($\downarrow$). }  
    \label{tab:sw_cmp}
        \begin{tabularx}{0.944\linewidth}{  c || c || c | c | c || c | c }
            \hline   
            Method & Avg $L_1$ & Prec. & Recall & $F_1$ & Avg D. & Max D. \\

            \hline
            \rowcolor{OurTableColor}
            BBW & 0.575 & 45.80 & 97.06 & 56.58 & 0.009 & \textbf{0.410} \\%
            GV & 0.687 & 42.14 &  \textbf{99.99} & 55.54  & 0.009 & 0.658 \\
            \rowcolor{OurTableColor}
            GV* & 0.645 & 75.11 & 89.24 & 76.80 & 0.008 & 0.7046 \\
            RigNet & 0.153 & \textbf{98.11} & 87.17 & 90.41 & \textbf{0.002} & 0.697 \\
            \rowcolor{OurTableColor}
            Ours & \textbf{0.124} & 88.70 & 98.78 & \textbf{91.72} & \textbf{0.002} & 0.692 \\
            Ours-S & 0.135 & 98.09 & 88.85 & 91.53 & \textbf{0.002} & 0.697  \\%
            \hline   
    \end{tabularx}
    \vspace{-.1in}
    }
\end{table}

\subsubsection{Ablation} 
In Figure~\ref{fig:sw_ablation} we modify different parts of our algorithm for this application. For each option we show the predicted weights and the $L_1$ error. We use harmonic weights instead of MVC, which results in a minor increase in the $L_1$ error (third column). We remove the $L_p$ loss leading to a significant increase in the $L_1$ error, as well as smoothing of the error, and its distribution in a larger area, as expected (fourth column). Finally, we remove the symmetry loss leading again to an increase in the $L_1$ loss, as well as loss of symmetry of the weights, which can be seen by the loss of symmetry of the $L_1$ error function (fifth column).

Additionally, Table \ref{tab:sw_ablation}, shows the quantitative ablation results on the Roblox dataset. We note that the full configuration, with MVC and all the loss parts, achieves the best average $L_1$ loss, closely followed by the harmonic coordinates (as in Figure~\ref{fig:sw_ablation}). We conjecture that even though the dataset has interior parts, for which harmonic coordinates do not work as well), for skinning weights we expect the interior parts to deform similarly to the exterior parts, thus allowing for a good weight approximation by harmonic coordinates. Naturally, these are much more expensive to compute, so we provide this example for illustrative purposes only. The $F_1$ score is similar for all configurations, with the exception of removing the $L_p$ loss. In this case, the precision reduces significantly, as we falsely identify many more bones as influential (since the weights are not as sparse), leading to more false positives. The average and maximal vertex distance error on the animated frames is also similar between the different configurations. The smallest maximal distance is obtained by removing the $L_p$ loss and reducing sparsity, leading to a more uniform error distribution and smaller maximal error.

\begin{table}[b]
    \centering
    {\smaller
    \caption{We show an ablation study with different cage coordinates and loss terms, using the same metrics as in Table~\ref{tab:sw_cmp}. See the text for details.}
    \label{tab:sw_ablation}
        \begin{tabularx}{0.934\linewidth}{  c || c || c | c | c || c | c }
            \hline   
            Ablation & Avg $L_1$ & Prec. & Recall & $F_1$ & Avg D. & Max D. \\

            \hline       
            \rowcolor{OurTableColor}
            CageNet & \textbf{0.124} & 88.70 & 98.78 & 91.72 & \textbf{0.002} & 0.692 \\
            Harmonic & 0.127 & 89.21 & 98.54 & 91.92 & \textbf{0.002} & 0.705 \\
            \rowcolor{OurTableColor}
            $\lambda_p=0$ & 0.135 & 59.10 & \textbf{99.93} & 71.15 & \textbf{0.002} & \textbf{0.688} \\
            $\lambda_{\text{sym}}=0$ & 0.133 & \textbf{89.47} & 98.49 & \textbf{92.06} & \textbf{0.002} & 0.748 \\
            \hline   
    \end{tabularx}
    \vspace{-.1in}
    }
\end{table}

\section{Limitations, Future Work and Conclusion}
\begin{figure}[t]
    \centering
    \includegraphics[width=\linewidth]{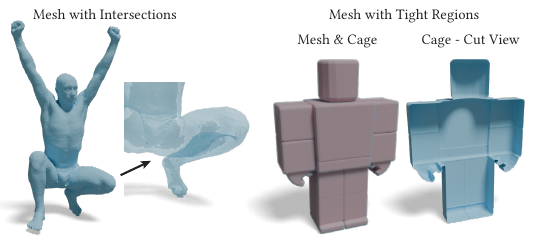}
    \caption{Limitations. CageNet learns parametrized volumetric \emph{functions}, and thus it cannot represent multiple values at the same spatial position. Thus, intersections in the input mesh (left) will be problematic if labeled differently. Additionally, if the mesh has large areas in close proximity (right), the cage may envelope them together, restricting the diversity of functions representable in these areas.}
    \label{fig:limitations}
\end{figure}

\begin{figure*}
    \centering
    \includegraphics[width=\linewidth]{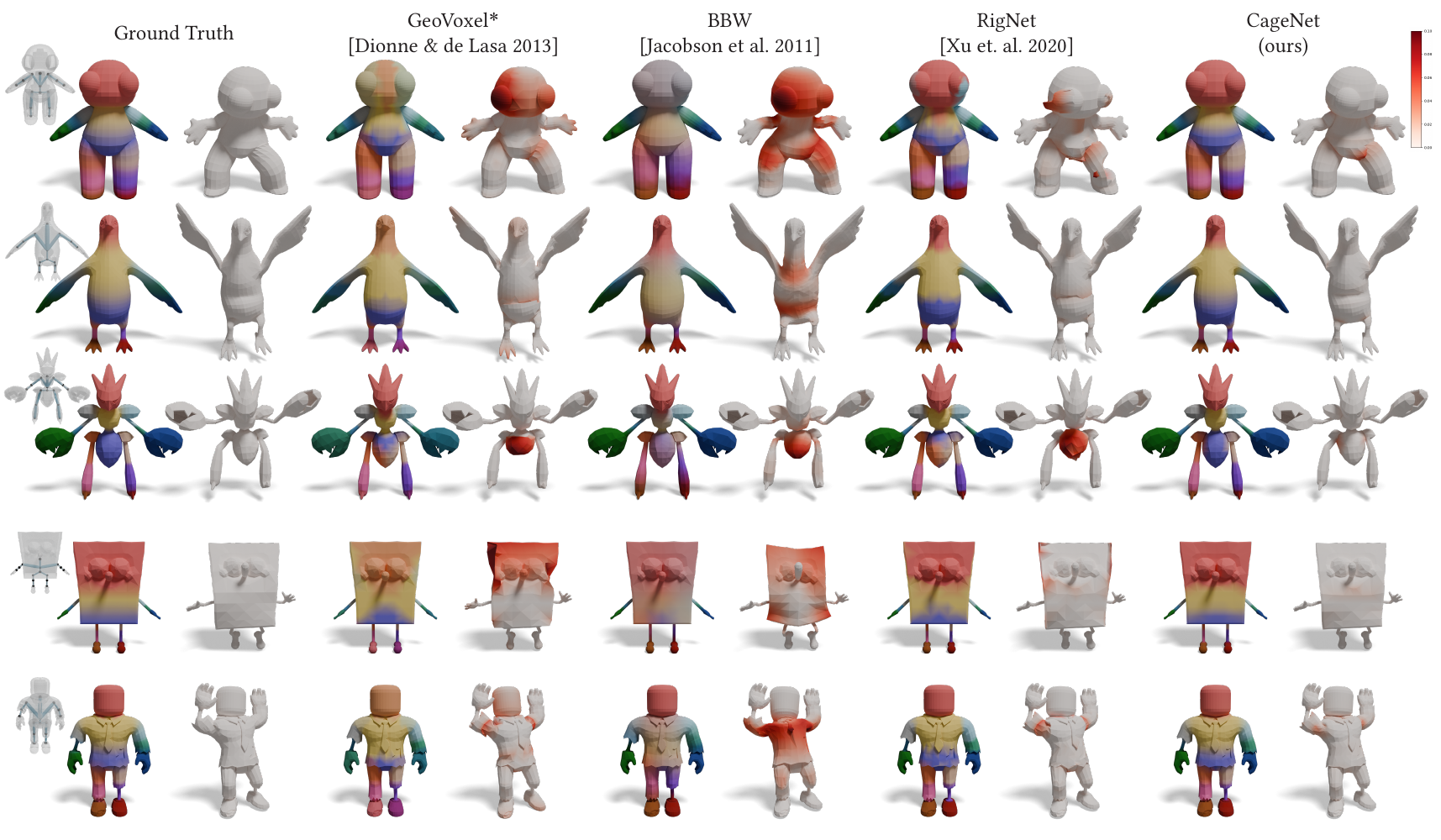}
    \caption{Comparison against baselines, each column shows the skinning weights (left) and an animation frame resulting from these weights, color coded with the corresponding normalized vertex displacement error (right). Here GeoVoxel* uses the parameters recommended by \citet{XuZKLS20}. Our approach achieves better skinning weight predictions compared to the baselines, leading to less artifacts in the resulting animation frame. 
    Please refer to the supplementary video for the full animation sequences. }
    \label{fig:sw_cmp_test}
\end{figure*}

\begin{figure*}
    \centering
    \includegraphics[width=\linewidth]{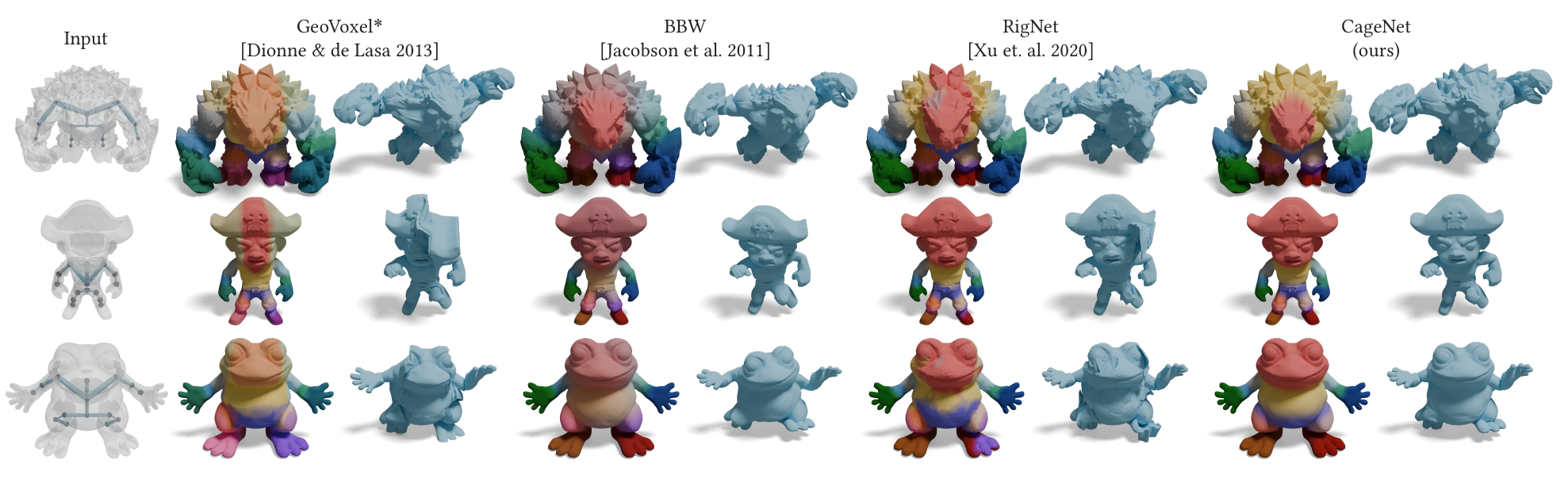}
    \caption{Our method can be applied to AI-generated meshes, such as \cite{xiang2024structured}. We show qualitative comparisons of the generated skinning weights (left) and a resulting deformation frame (right) on these characters. Here GeoVoxel* uses the parameters recommended by \citet{XuZKLS20}. As the ground truth weights are not available here, compare with the dataset ground truth in Figure~\ref{fig:sw_cmp_test}. We note that our approach leads to weights which are most similar to the ground truth, with the least artifacts in the deformed frame. Please refer to the supplementary video for the full animation sequences.  }
    \label{fig:sw_cmp_ai}
\end{figure*}

Meshes that have self-intersections (Figure~\ref{fig:limitations} left) are problematic for CageNet if the regions that intersect have different labels or feature values. Since our features are volumetric scalar \emph{functions} on the interior of the cage, they cannot be multi-valued at a given point. Using multi-valued functions (e.g., polynomial roots represented by the polynomial coefficients) may alleviate this restriction in future work. 
Despite the cage-offset augmentation (Figure~\ref{fig:seg_cmp_topo}), our approach is still somewhat sensitive to topological inconsistencies due to cage generation (such as the extreme case in Figure~\ref{fig:limitations} right). Future improvements on cage generation, such as adaptive offsets or a topologically robust method (e.g., \cite{zint2024topological}), could further improve the stability of CageNet.
We demonstrate the efficacy of our CageNet using DiffusionNet~\cite{SharpACO22} as an example. Future explorations on combining our approach with different network architectures and with other generalized barycentric coordinates (e.g., \cite{ThieryMC24,GoesD24}) could further boost CageNet's performance, and improve its efficacy in  applications (e.g., generalizing to different skeleton topologies). 

To conclude, we have shown that caging wild meshes is beneficial for data-driven shape processing with neural networks. We believe that CageNet can find many additional applications, and inspire future work in the realm of robust geometry processing.

\begin{acks}
Mirela Ben Chen acknowledges the support of the Israel Science Foundation (grant No. 1073/21).
We thank Weiqi Shi for collecting and providing the Roblox dataset. We also use models from Turbosquid, the Windows 3D Library, and Sketchfab, and we thank the respective creators for making their models publicly available. We acknowledge the Blender community for making their software publicly available~\cite{blender,Liu_BlenderToolbox_2018}.
\end{acks}

\bibliographystyle{ACM-Reference-Format}
\bibliography{cagenet_bib}

\end{document}


\title{CageNet: A Meta-Framework for Learning on Wild Meshes \\ Supplemental Material}


\author{Michal Edelstein}
\orcid{0000-0001-9126-1617}
\affiliation{%
    \institution{Technion - Israel Institute of Technology}
    \city{Haifa}
    \country{Israel}}
\email{edel.michal@gmail.com}

\author{Hsueh-Ti Derek Liu} 
\orcid{0009-0001-1753-4485}
\affiliation{%
    \institution{Roblox \& University of British Columbia}
    \city{Vancouver}
    \country{Canada}}
\email{hsuehtil@gmail.com}

\author{Mirela Ben-Chen} 
\orcid{0000-0002-1732-2327}
\affiliation{%
    \institution{Technion - Israel Institute of Technology}
    \city{Haifa}
    \country{Israel}}
\email{mirela@cs.technion.ac.il}


\acmSubmissionID{491}

\ccsdesc[500]{Computing methodologies~Shape analysis}

\maketitle

\section{Hyper Parameters}

For both segmentation and skinning weights prediction, we use the basic 4-block DiffusionNet architecture \cite{SharpACO22}, with spectral acceleration and $k=128$ eigenbasis. We vary the width between the applications, where we use a 64-width NN for the segmentation and a 128-width NN for the skinning weights.
We use an ADAM optimizer with an initial initial learning rate of $0.0005$ for segmentation and $0.001$ for skinning weights. For both applications, we use a batch size of 1, and train for 200 epochs while decaying the learning rate by a factor of 0.5 every 50 epochs, as suggested by \citet{SharpACO22}.
We also normalize the cages to the unit sphere, and, when HKS features are used, we utilize DiffusionNet's computation and default parameters -- the heat kernel signatures are sampled at $16$ values of $t$ logarithmically spaced on [0.01, 1].

\section{Skinning Weights Dataset}

\begin{figure*}
    \centering
    \includegraphics[width=\textwidth]{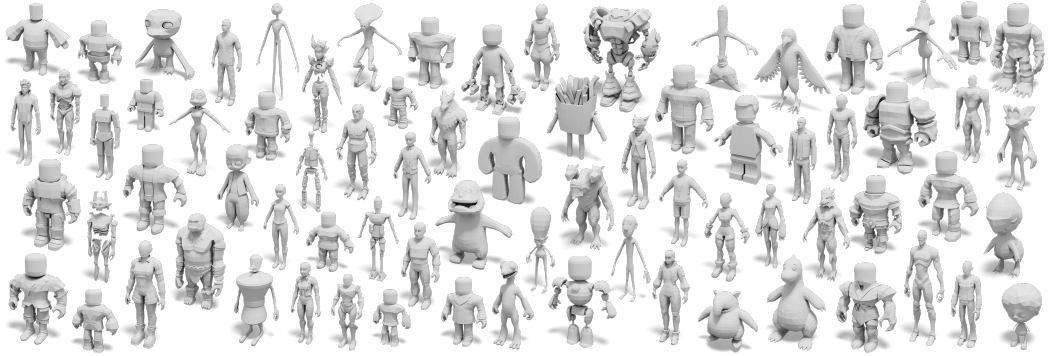}
    \caption{To evaluate our method on in-the-wild meshes, we collected 753 artist-created characters from the Roblox platform. Here we present samples from the dataset to demonstrate the variety of characters with different complexities.}
    \label{fig:roblox_dataset} 
\end{figure*}

We evaluate the performance on skinning weights generation on a dataset of 753 artist-created biped characters (see Figure ~\ref{fig:roblox_dataset}) obtained from the Roblox platform, where we hold out 75 meshes as the test set. Meshes in the dataset share the same pose, orientation and skeleton topology. 

\section{Skinning Weights Symmetry Loss}
Let $s \in \xR^{n\times k}$ be the ground truth skinning weights matrix, and take $s_i$ to be its $i$-th column. Further, let $\{\mathcal{S}\}_{i=1}^{N}$ be the set of symmetric vertices, i.e. those which have a close match when reflected around the $yz$ plane. Define $A \in {\{0,1\}}^{n\times n}$ as the diagonal binary matrix that extracts the symmetric vertices, namely $A_{ii} = 1$ iff $v_i \in \mathcal{S}$. Hence, for a vector $h$, $Ah$ zeros out all the entries of the non-symmetric vertices. We use $A$ to identify all the joints whose weights have a \emph{large support on symmetric vertices}, by finding all indices $1 \leq p \leq k$ such that $||A s_p ||_1 > \delta_s$. Denote this set by $\mathcal{I}$.

Now, take $B \in {\{0,1\}}^{n\times n}$ to be the binary matrix that maps the symmetric vertices to their match. Thus, $B_{ij} = 1$ iff $v_i,v_j$ are symmetric w.r.t. $yz$ plane. For each $p\in \mathcal{I}$ we check if there exists a joint $q$, such that the weights $s_p,s_q$ are symmetric on the symmetric vertices. Specifically, for each $p \in \mathcal{I}$ we find $1 \leq q \leq k$ such that $\frac{\|A s_p - B s_q \|}{N\|s_p\|} < \epsilon_s$. We take $\epsilon_s = 10^{-5}, \delta_s=30$.

Finally, let $\mathcal{K} = \{(p,q)\}$ be the set of all such pairs. Compute for each pair $l=(p,q)$ and each vertex $1 \leq i \leq n$ the symmetry error $w(i,l) = \big((A \hat{s}_p)(i) - (B \hat{s}_q)(i)\big)^2 $, where $\hat{s}$ are the predicted skinning weights. The symmetry loss is given by:
\begin{equation*}
Sym(\hat{s},s) = \frac{1}{n} \sum_{i=1}^n \sqrt{\sum_{l=1}^{|\mathcal{K}|} w(i,l)}.
\end{equation*}

\section{\ME{CageNet vs. Mesh Repair}}
\begin{figure}
    \centering
    \includegraphics[width=\linewidth]{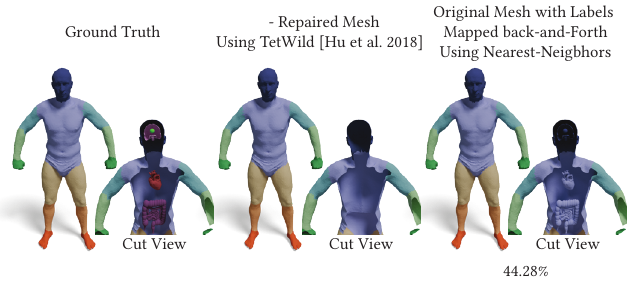}
    \caption{\ME{For a mesh with internal components (left), repairing it using TetWild~\cite{Hu2018tetwild} yields a repaired surface (center), such that mapping the labels back and forth from the input to the repaired surface and back leads to a high labeling error (right). Compare with Fig. 5 in the main paper.}}
    \label{fig:internal-fix}
\end{figure}

\ME{
Repairing the meshes and then applying a baseline network is problematic for two main reasons. First, meshes may have many variants of issues (non manifold elements, internal components, multiple components), and each such issue will require a different repair method, the choice of which is sometimes manual (and thus not scalable to large datasets). Second, one would still need to map the features between the input mesh and the repaired mesh. Such mapping (e.g. using nearest neighbors) introduces large errors in the functions. For example, for the experiment in Figure 5 in the paper, if we repair the mesh using TetWild~\cite{Hu2018tetwild} (obtaining only the exterior surface), and then map the presented function back and forth, we get low accuracy $(44.28\%)$ since the internal components are mapped to the exterior surface. Using a cage and barycentric coordinates solves this issue (accuracy $99.98\%$ as we show in Fig. 5), since our setting provides an accurate representation of volumetric functions using a manifold surface. Figure~\ref{fig:internal-fix} demonstrates this.
}

\section{\ME{Segmentation - Additional Results}}

\ME{
In Figure \ref{fig:seg_ours_additional}, we test our segmentation network on wild meshes containing multiple connected components, including a non-manifold one. The segmentation results are consistent with the ground truth of a training mesh shown for reference.
}

\section{\ME{Skinning Weights - Additional Results}}

\begin{figure*}
    \centering
    \includegraphics[width=\textwidth]{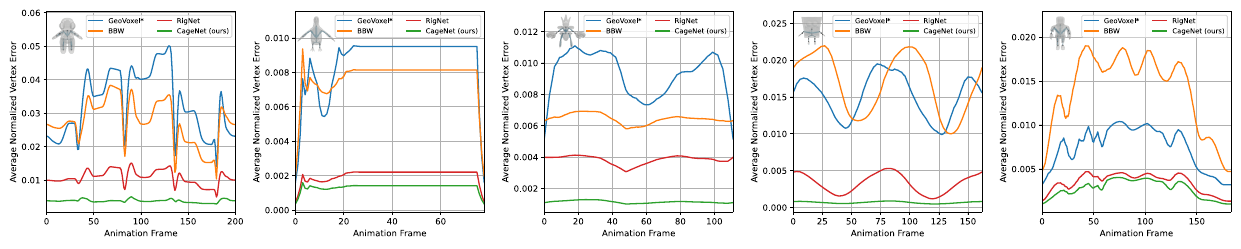}
    \caption{\ME{The average normalized vertex error per frame for all the methods and meshes displayed in the video. Our method has the lowest error across the entire video.}}
    \label{fig:sw_vid_err} 
\end{figure*}

\ME{\paragraph{Quantitative Comparison} Figure \ref{fig:sw_vid_err} shows the average normalized vertex displacement error for each of the animations presented in the Supplemental video. Note that for all $5$ animation sequences, our mean error is lower than the other methods.}

\begin{figure*}
    \centering
    \includegraphics[width=\textwidth]{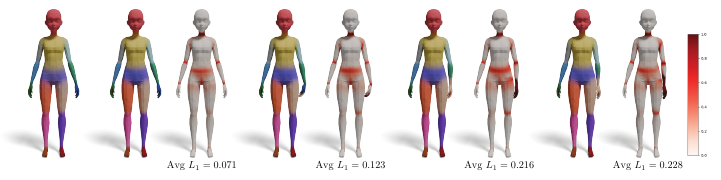}
    \caption{\ME{To demonstrate the effect of topological changes on our skinning results, we deformed a mesh from the test set by moving its arm closer to the body. As the arm approaches, the cage encompasses a larger area of the hand and thigh together, resulting in a larger error in this region.}}
    \label{fig:sw_topo} 
\end{figure*}

\begin{figure*}
    \centering
    \includegraphics[width=0.85\textwidth]{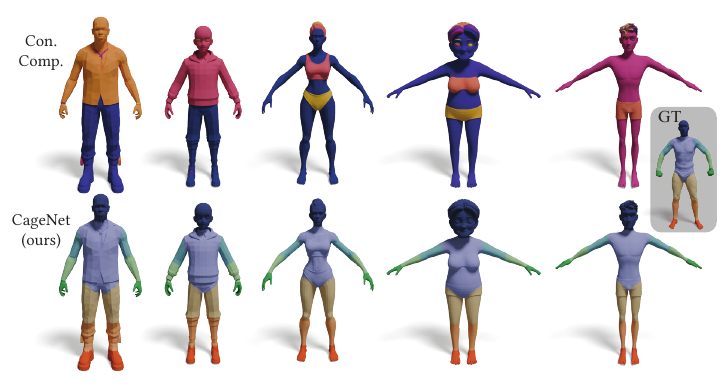}
    \caption{\ME{Segmentation results on wild meshes. Each mesh contains multiple connected components, visualized using distinct colors (first row), with the corresponding segmentation results shown in the second row. The meshes are taken from the Roblox dataset and from Turbosquid \cite{turbosquid}}}
    \label{fig:seg_ours_additional} 
\end{figure*}

\ME{\paragraph{Topological Changes} We show our method's performance given topological changes in the cage in Figure \ref{fig:sw_topo}. Notably, since our training dataset consists of shapes in the A-pose, the neural network was not trained on cages with such topological variations, or the cage offset augmentation, which were unnecessary for this dataset. There, it can be that as the hand grows closer to the body, the error increases. To achieve improved robustness to such variations, the training set would need to incorporate meshes with diverse poses and training should be done with the cage offset augmentation.}

\begin{figure*}
    \centering
    \includegraphics[width=\textwidth]{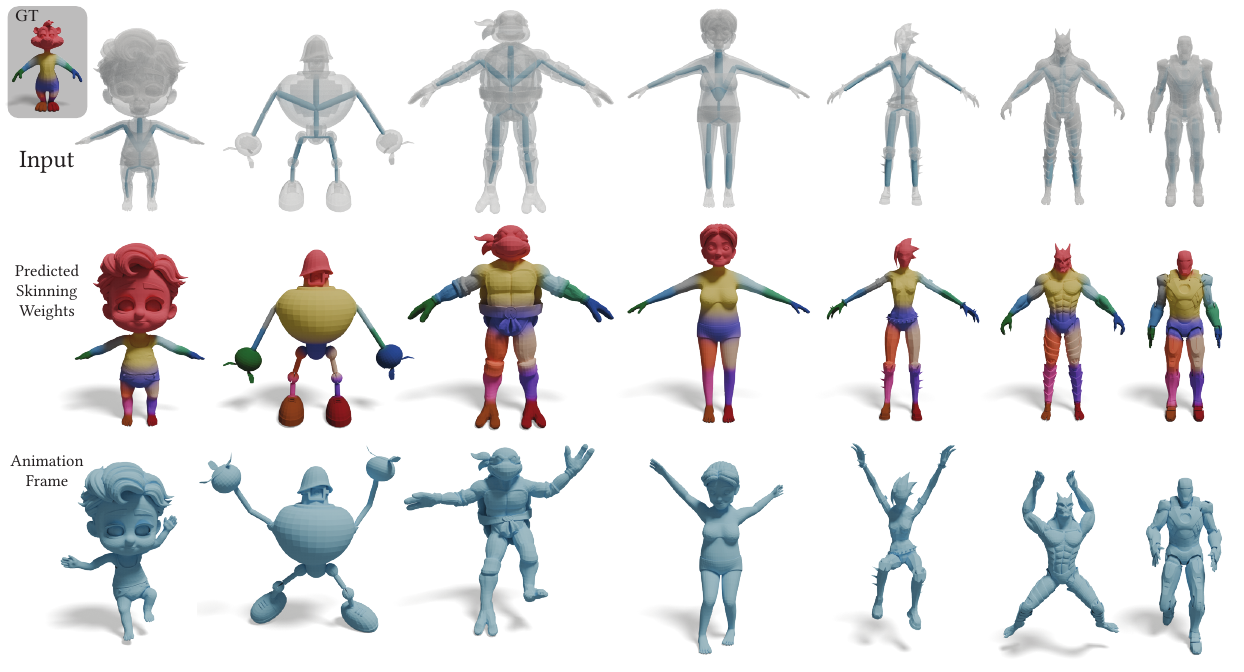}
    \caption{\ME{Skinning weights results on wild meshes containing multiple connected components and non-manifold elements. For each model (top row), we manually created a skeleton consistent with our dataset. We then applied our skinning weight prediction network and show the resulting weights (middle row), as well as one animation frame per model (bottom row). Model sources: Turbosquid \cite{turbosquid}, Windows 3D Library \cite{windows3dmodels}, Sketchfab \cite{sketchfab, Raphael, Mania, Kaiju}.
    }}
    \label{fig:sw_additional_rebt} 
\end{figure*}

\ME{
\paragraph{More Wild Meshes} We also show additional results on wild meshes in Figure \ref{fig:sw_additional_rebt}. A skeleton was manually created for each shape. The predicted skinning weights align well with the ground truth weights, shown on a reference model from the Roblox dataset. For each model, we include an animation frame to illustrate the effectiveness of the predicted weights.
}

\bibliographystyle{ACM-Reference-Format}
\bibliography{cagenet_bib}